\documentclass[reprint]{revtex4-1}

\usepackage[utf8]{inputenc}
\usepackage[T1]{fontenc}	
\usepackage[english]{babel}
\usepackage{amsmath,amssymb,amsfonts, bm,bbm,slashed, subdepth}
\usepackage{graphicx}
\usepackage[sort&compress]{natbib}
\usepackage{xcolor}
\usepackage[normalem]{ulem}
\usepackage{hyperref}
\usepackage{cleveref}
\usepackage{enumerate}
\usepackage{epsfig, subfigure}
\usepackage{booktabs, tabularx}
\usepackage{units}	

	\usepackage{amsmath}
	\allowdisplaybreaks[1]						
	\usepackage{amssymb}
	\usepackage{amsfonts}						
	\usepackage{braket}						
	\usepackage{slashed}						
	\usepackage{bbold}
	\usepackage{upgreek}
	\usepackage{multirow}
	\usepackage{nicefrac}
	\usepackage{xfrac}
	{

\begin{document}

\title{Higgs boson results on couplings to fermions, CP parameters and perspectives for HL-LHC (ATLAS AND CMS) \\ \vspace{3mm}\normalsize Talk presented at the International Workshop on Future Linear Colliders (LCWS2017),\\ Strasbourg, France, 23-27 October 2017. C17-10-23.2.}
\author{Johannes Brandstetter\footnotetext[*]{This is my footnote}}
\email{Institute of High Energy Physics, Austrian Academy of Sciences \\ johannes.brandstetter@cern.ch}
\affiliation{on behalf of the ATLAS and CMS Collaborations}

\begin{abstract}
This report summarizes latest ATLAS and CMS results on Higgs boson couplings to fermions. Presented topics include decays into final states of pairs of tau leptons and pairs of bottom quarks as well as results on the ttH production mode. Results are complemented by tests of the CP invariance and searches for lepton flavor violating decays.~Finally, prospects of future Higgs boson analyses within the scope of the High Luminosity LHC program are discussed. The presented results mostly use LHC 2016 data collected at a center-of-mass energy of $\sqrt{\mathrm{s}}=13~$TeV corresponding to an integrated luminosity of about 36~$\mathrm{fb^{-1}}$.
\end{abstract}
	
\maketitle

\section{Introduction}
In the standard model (SM), Yukawa couplings describe interactions of fermions with the Higgs boson field.~Since Yukawa couplings are proportional to the mass of the fermions the best way to probe Yukawa couplings are in Higgs boson decays into pairs of tau leptons and pairs of bottom quarks.~Couplings to top quark pairs can be probed via the gluon fusion or the associated ttH production mode.~Any deviations of Yukawa couplings, e.g. asymmetries in up/down quarks or lepton flavor violating Higgs boson decays, would be strong signs of new physics.

\section{Results on Higgs boson couplings to fermions}
\subsection{Decay into pairs of tau leptons}
\begin{figure}[htb]
\begin{center}
\includegraphics[width=0.5\textwidth]{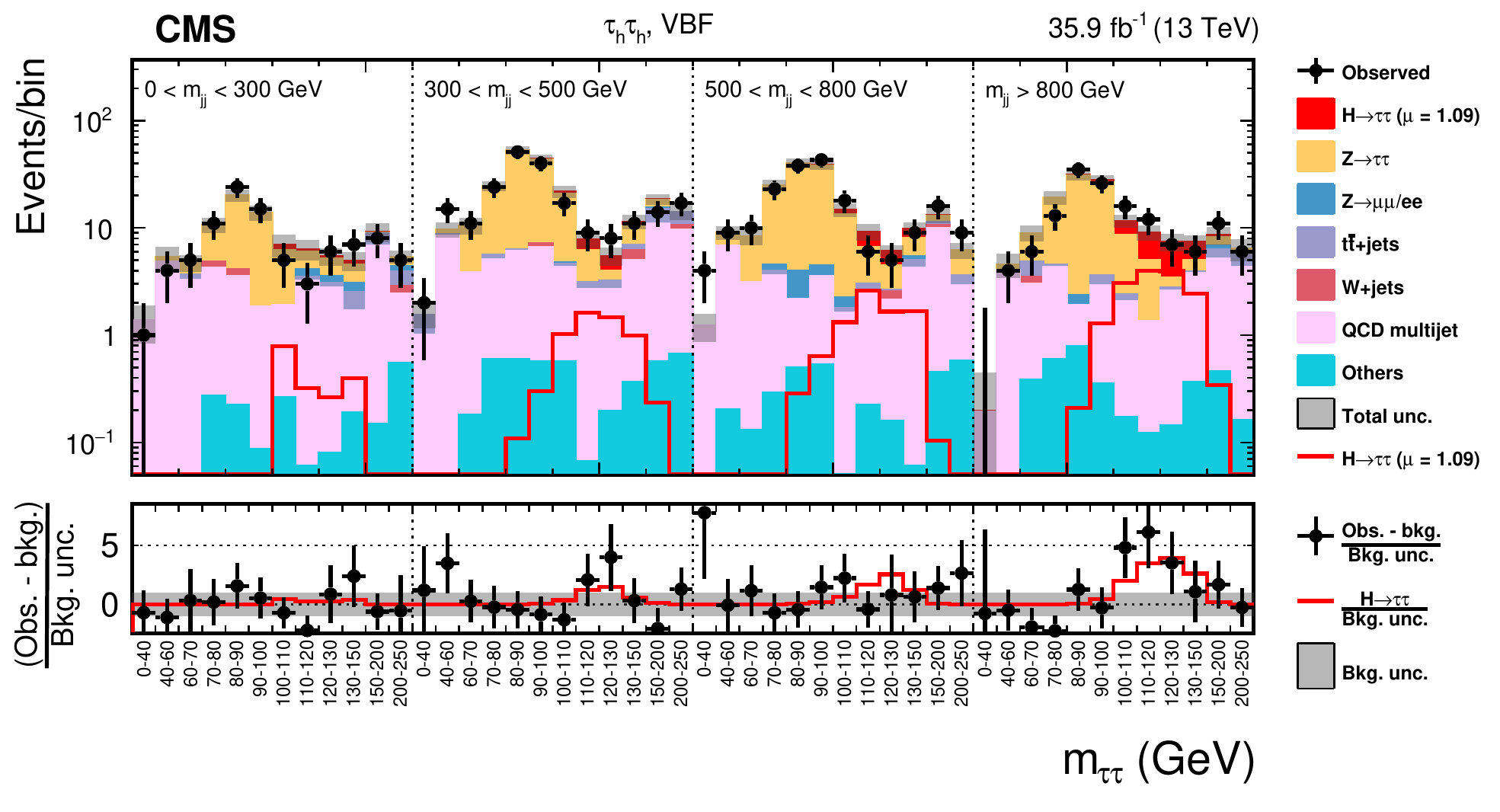}
\caption{Observed and predicted 2D distributions in the 2-jet VBF category of the $\uptau_{\text{h}}\uptau_{\text{h}}$ decay channel \cite{CMS_Htautau_Run2}.~In this category the 2D distribution consists of the mass of the di-tau system ($\mathrm{m_{\uptau\uptau}}$) and the mass of the leading and subleading jet ($\mathrm{m_{jj}}$).  }
\label{fig:2D_VBF}
\end{center}
\end{figure}

The larger mass and therefore the larger coupling with respect to the di-muon final state, and the smaller quark and gluon jet background, and therefore the better experimental accessibility with respect to the b quark final state, make Higgs boson decays into pairs of tau leptons to most promising candidate to probe Yukawa couplings.~Previous analyses that use data collected at $\sqrt{\mathrm{s}}=7~$TeV and $\sqrt{\mathrm{s}}=8~$TeV (the data taking period in 2011 and 2012 is from now on referred to as LHC Run-1) can be found in Refs.~\cite{ATLAS_Htautau_Run1,CMS_Htautau_Run1} for ATLAS~\cite{ATLAS_detector} and CMS~\cite{CMS_detector}, respectively.~The measurements are compatible with each other and with SM predictions.~The combination of the two individual measurements \cite{ATLAS_CMS_Run1_combination} establishes the Higgs boson decay into pairs of tau leptons for the first time with a significance larger than five standard deviations over the background-only hypothesis. \\ \\
The first direct observation of the Higgs boson decay into pairs of tau leptons is reported by CMS in Ref.\cite{CMS_Htautau_Run2}.~Since tau leptons can both decay into hadrons and leptons, the analysis has to deal with several final states.~The most sensitive final states comprise $\uptau_{\text{h}}\uptau_{\text{h}}$, $\uptau_{\text{h}}\uptau_{\mu}$, $\uptau_{\text{h}}\uptau_{\mathrm{e}}$ and $\uptau_{\mu}\uptau_{\mathrm{e}}$ decay modes, where $\uptau_{\text{h}}$ denotes a hadronically decaying tau lepton and $\uptau_{\mu}$ ($\uptau_{\mathrm{e}}$) a tau lepton decaying into a muon (electron). \\ The analysis is split into three mutual exclusive event categories, namely a 0-jet category to target Higgs boson production via gluon fusion, a 1-jet boosted category, and a 2-jet category to target Higgs boson production via vector boson fusion (VBF).~The signal is extracted from simultaneous maximum likelihood fits in 2D categories that are chosen to maximize the discovery potential.~The event distribution in the most sensitive category, the 2-jet VBF category, is shown in Fig.~\ref{fig:2D_VBF}.

\begin{figure}[htb]
\begin{center}
\includegraphics[width=0.5\textwidth]{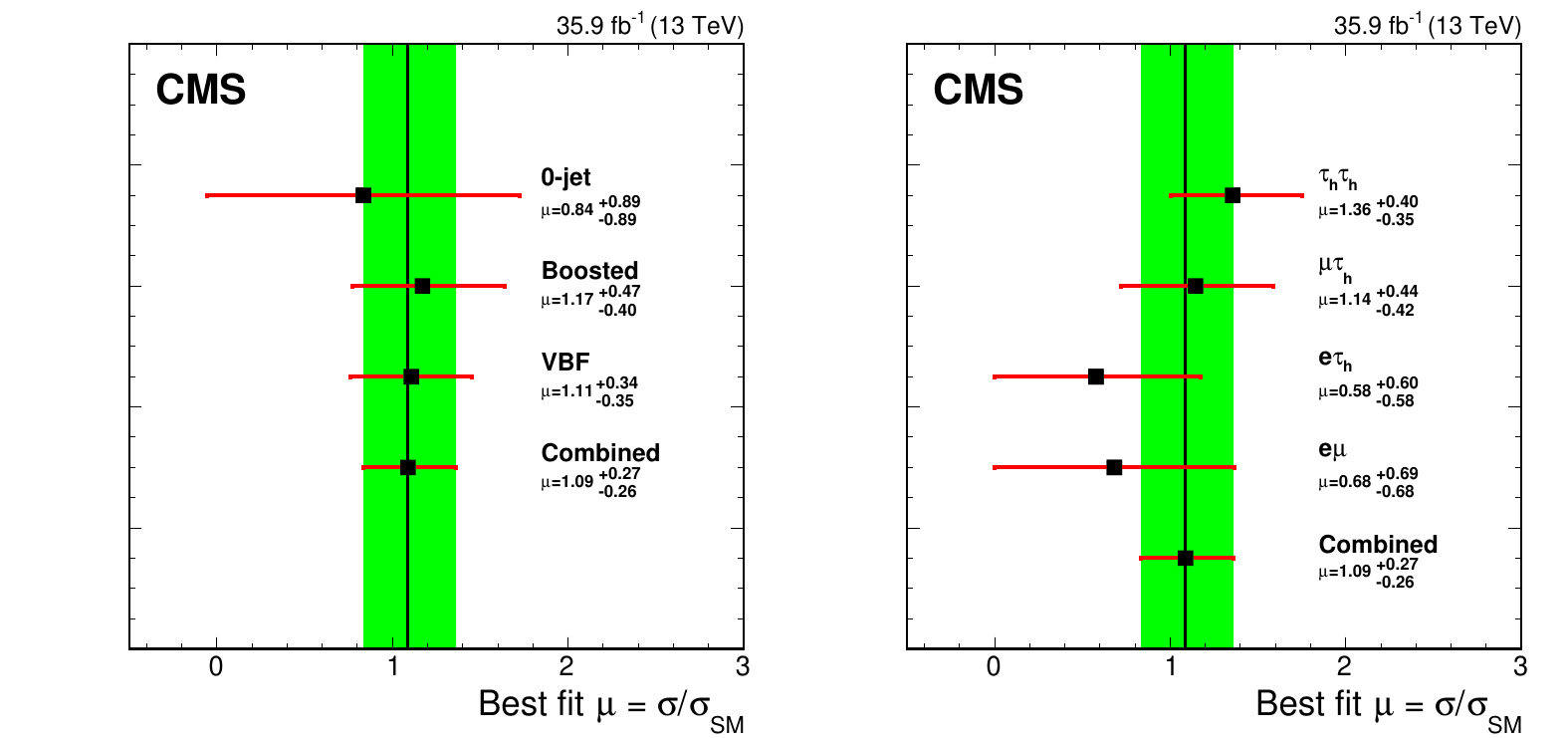}
\caption{Best fit signal strength per category (left) and channel (right) \cite{CMS_Htautau_Run2}. }
\label{fig:Htautau_signalStrength}
\end{center}
\end{figure}

The analysis reveals an excess in data where the observed (expected) significance for a SM Higgs boson with $\mathrm{m}_\text{H}=125.09~$GeV is 4.9 (4.7) standard deviations.~The corresponding best fit value for the signal strength is found to be $1.09^{+0.27}_{-0.26}$ at $\mathrm{m}_{\text{H}}=125.09~$GeV.~Here, $\mathrm{m}_{\text{H}}=125.09~$GeV corresponds to the combined ATLAS and CMS Run-1 measurement on the Higgs boson mass~\cite{Higgs_mass}.~The best fit values for the signal strengths per channel and per category are shown in Fig.~\ref{fig:Htautau_signalStrength}. \\
Results are combined with the results of the CMS LHC Run-1 search for a Higgs boson decaying into pairs of tau leptons \cite{CMS_Htautau_Run1} yielding a combined observed (expected) significance of 5.9 (5.9) standard deviations.

\subsection{Decay into pairs of bottom quarks}

\begin{figure}[htb]
\begin{center}
\includegraphics[width=0.4\textwidth]{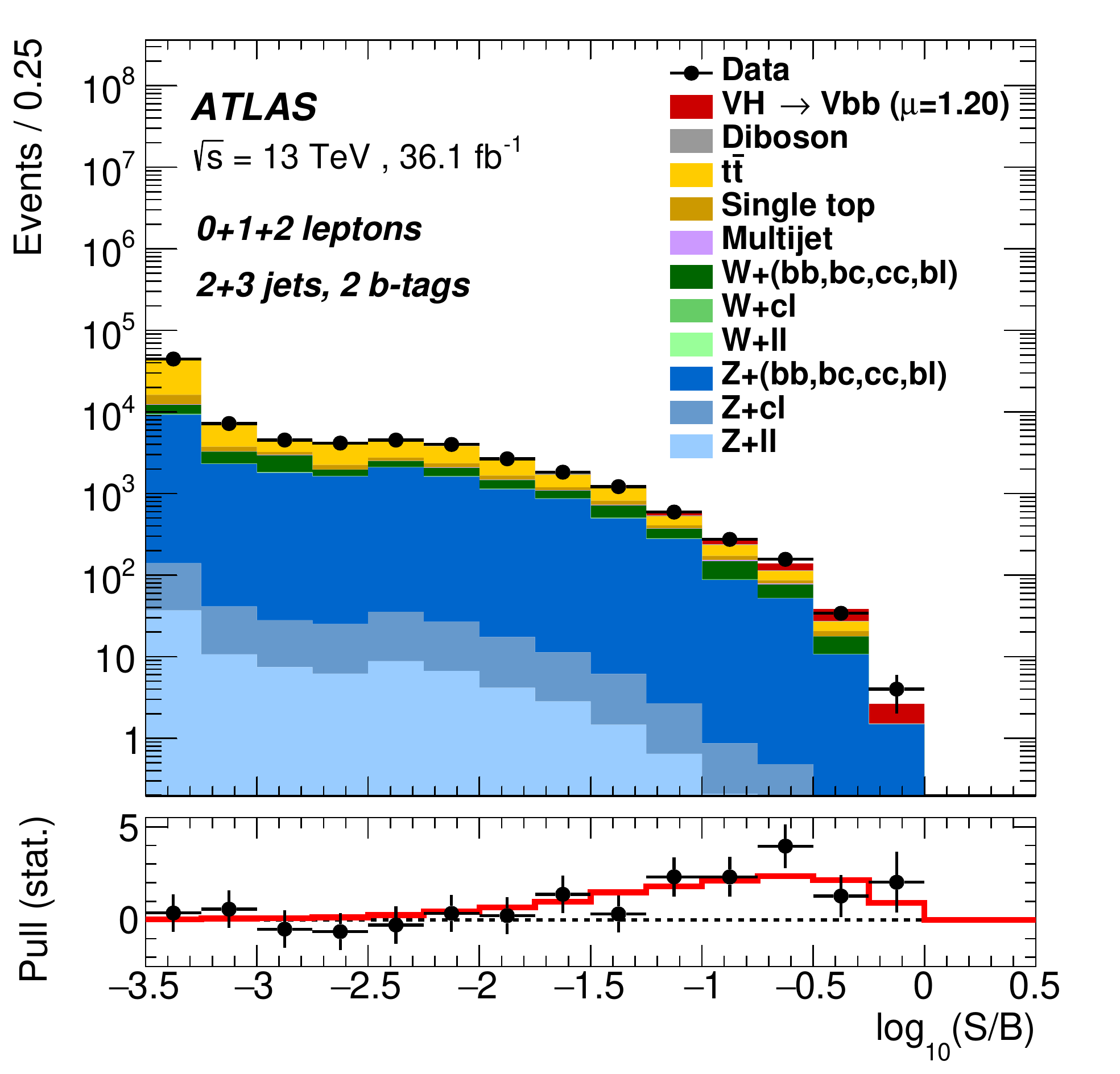}
\caption{ Event yield as a function of log(S/B) for data, background and a Higgs boson with $\mathrm{m}_{\text{H}}=125~$GeV \cite{ATLAS_Hbb_Run2}. }
\label{BDT_output_ATLAS}
\end{center}
\end{figure}

In contrast to the di-tau final state, the gluon fusion and VBF production modes are less sensitive for searches of Higgs boson decays into pairs of bottom quarks due to the overwhelming multi-jet backgrounds.~Therefore, most analyses rather target Higgs boson production in association with W or Z bosons.~Previous bb final state analyses comprise results from CDF and D0 experiments at Tevatron as well as ATLAS and CMS results~\cite{Tevatron_Hbb,ATLAS_Hbb_Run1,CMS_Hbb_Run1,ATLAS_CMS_Run1_combination}. \\ \\
The most recent ATLAS result can be found in Ref.~\cite{ATLAS_Hbb_Run2}.~Channels are distinguished by the number of charged leptons (0-lepton, 1-lepton, 2-lepton) in the final state in addition to two b-tagged jets.~The 0-lepton category targets decays of the Z boson into two neutrinos, the 1-lepton category decays of the W boson into lepton plus neutrino and the 2-lepton category decays into two leptons.~Further subcategorization is made concerning the transverse momentum of the reconstructed vector boson and the number of jets.~A multivariate analysis strategy is applied where the output of Boosted Decision Trees (BDTs) is used as  discriminating variable.~Two cross-check analyses are performed: The first applies a maximum likelihood fit to the bb mass system, the second uses the same BDT strategy but is trained on (W,Z)Z resonances where the Z boson decays into pairs of bottom quarks.~Figure~\ref{BDT_output_ATLAS} shows the result of the HZ analyses where the final-discriminant bins are combined into bins of log(S/B). \\

The analysis reveals an excess in data where the observed (expected) significance for a SM Higgs boson is 3.5 (3.0) standard deviations.~This is in agreement with the significances obtained by applying a maximum likelihood fit to the bb mass system.~As a byproduct the (W,Z)Z resonance is measured with an observed (expected) significance of 5.8 (5.3) standard deviations with respect to the background-only hypothesis.~Combination with ATLAS LHC Run-1 results of a Higgs boson decaying into pairs of bottom quarks yield an observed (expected) significance of 3.6 (4.0) standard deviations.~The best fit value for the signal strength is found to be $1.2^{+0.42}_{-0.36}$ and $0.9^{+0.28}_{-0.26}$ for Run-2 and for Run-1 and Run-2 combinations, respectively. \\ \\
The corresponding CMS analysis~\cite{CMS_Hbb_Run2} is conceptually similar, also dividing the analysis into 3 channels depending on the number of charged leptons and deploying a multivariate analysis strategy (BDTs) where a cross-check analysis targets the (W,Z)Z resonance.~The observed (expected) significance for a SM Higgs boson amounts to 2.8 (3.3) standard deviations.~Combination with CMS Run-1 results yields an observed (expected) significance of 3.8 (3.8) standard deviations.~The best fit value for the signal strength is found to be $1.19^{+0.40}_{-0.38}$ and $1.06^{+0.31}_{-0.29}$ for Run-2 and for Run-1 and Run-2 combinations, respectively.

\subsection{Test of top quark couplings via the ttH production mode}

\begin{figure}[htb]
\begin{center}
\includegraphics[width=0.5\textwidth]{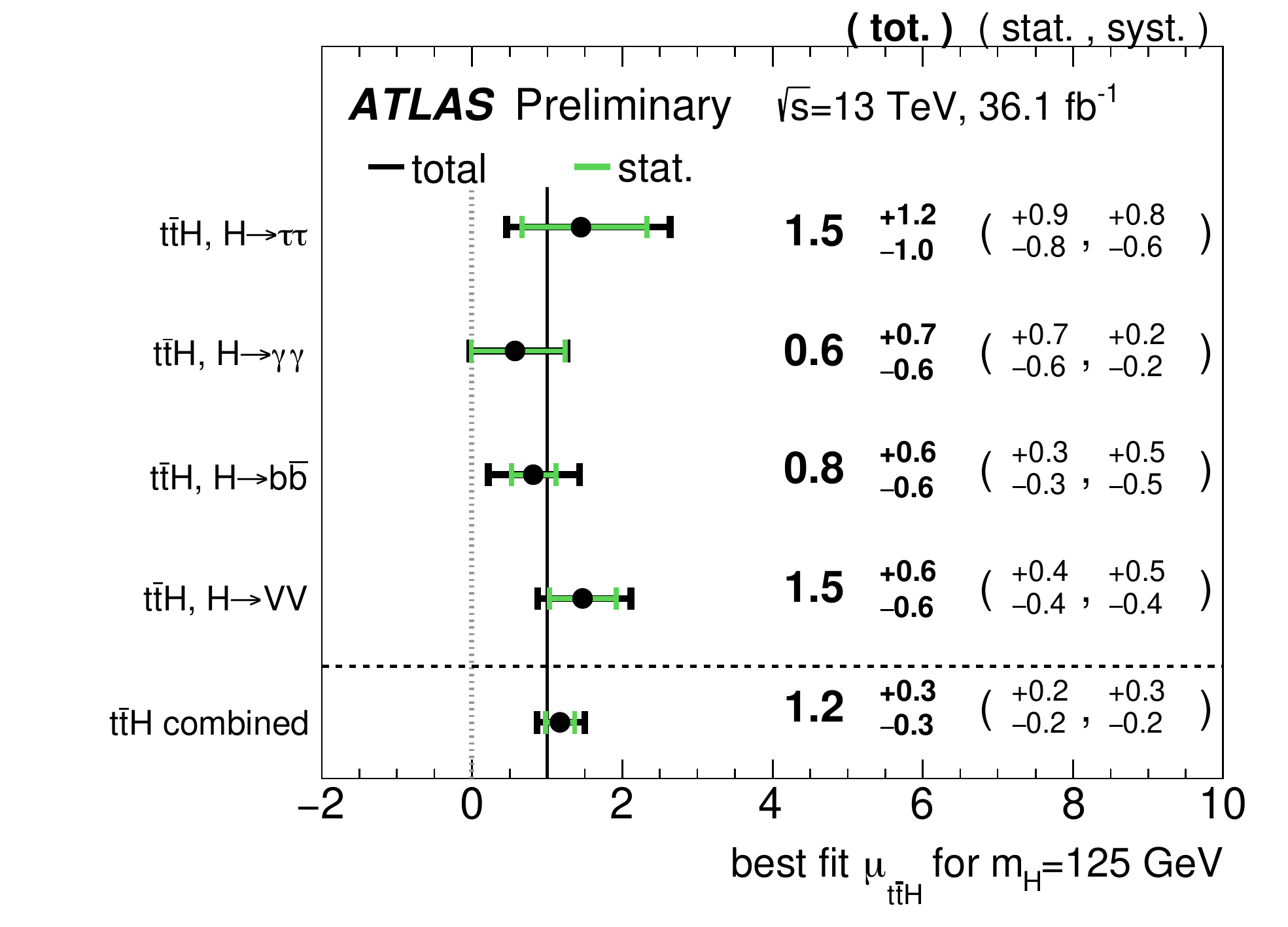}
\caption{ Best fit values of the signal strength $\mu_{\mathrm{ttH}}$ from individual results obtained by ATLAS \cite{ATLAS_ttH_comb_Run2}. }
\label{fig:ttH_combined_ATLAS}
\end{center}
\end{figure}

The Higgs boson production in association with a pair of top quarks is the most direct way to probe couplings of the top quark to the Higgs boson.~The production cross section is about 4 times higher at a center-of-mass energy of $\sqrt{\mathrm{s}}=13~$TeV than it is at $\sqrt{\mathrm{s}}=8~$TeV.~Combined LHC Run-1 results from ATLAS and CMS revealed a small excess of approximately two standard deviations with respect to the SM expectation that was most pronounced in the CMS same-sign lepton channel \cite{ATLAS_CMS_Run1_combination}.~However, no excess was seen in any other measurements where couplings of the top quark and the Higgs boson are involved, e.g. in gluon fusion tagged events or in decays of the Higgs boson into two photons.~Figure~\ref{fig:ttH_combined_ATLAS} summarizes the best fit values of the signal strength $\mu_{\mathrm{ttH}}$ from individual results obtained by ATLAS as well as the combined value that is found to be $\mathrm{\mu_{ttH} = 1.2^{+0.3}_{-0.3} }$.~Further ATLAS results on the ttH production mode using data collected at $\sqrt{\mathrm{s}}=13~$TeV cover combined results of Higgs boson decays into pairs of photons and pairs of Z bosons \cite{ATLAS_ttH_Run2} as well as decays into pairs of W bosons and pairs of tau leptons \cite{ATLAS_ttH_Run2_2}. \\ Recent CMS measurements of the ttH production mode and of the signal strength $\mu_{\mathrm{ttH}}$ are summarized in Tab.~I~\cite{CMS_ttH_Run2_1,CMS_ttH_Run2_2,CMS_ttH_Run2_3,CMS_ttH_Run2_4}.~Both ATLAS and CMS results are compatible with SM expectations but more data is needed to firstly, establish the ttH production with a significance of more than five standard deviations over the background-only hypothesis and secondly, to decrease the uncertainties on the measurements. 

\begin{center}
\begin{table}[htbp]
\begin{tabular}{ | c | c | c | c | }
\hline
\multirow{2}{*}{Channel} & \multirow{2}{*}{Dataset} & Obs. (exp.) & \multirow{2}{*}{$\displaystyle \mathrm{\frac{\mu_{ttH}}{\mu_{ttH}^{SM}}}$} \\
& & significance & \\ \hline \hline
$\gamma\gamma$ & 35.9~$\mathrm{fb^{-1}}$ & 3.3$\sigma$ (1.5$\sigma$) & $2.2^{+0.9}_{-0.8}$ \\ \hline
WW, ZZ, $\uptau\uptau$ & 35.9~$\mathrm{fb^{-1}}$ & 3.3$\sigma$ (2.5$\sigma$) & 1.5$^{+0.5}_{-0.4}$ \\ \hline
$\uptau\uptau$& 35.9~$\mathrm{fb^{-1}}$ & 1.4$\sigma$ (1.8$\sigma$) & 0.72$^{+0.62}_{-0.53}$ \\ \hline
bb & 12.9~$\mathrm{fb^{-1}}$ & - & -0.19$^{+0.80}_{-0.81}$ \\ \hline

\hline
\hline
\end{tabular}
\label{tab:ttH_combined_CMS}
\caption{ Summary of recent CMS measurements of the ttH production mode and the signal strength $\mu_{\mathrm{ttH}}$ \cite{CMS_ttH_Run2_1,CMS_ttH_Run2_2,CMS_ttH_Run2_3,CMS_ttH_Run2_4}. }
\end{table}
\end{center}

\section{Selected searches for BSM couplings}

\subsection{Lepton flavor violating decays}

\begin{figure}[htb]
\begin{center}
\includegraphics[width=0.4\textwidth]{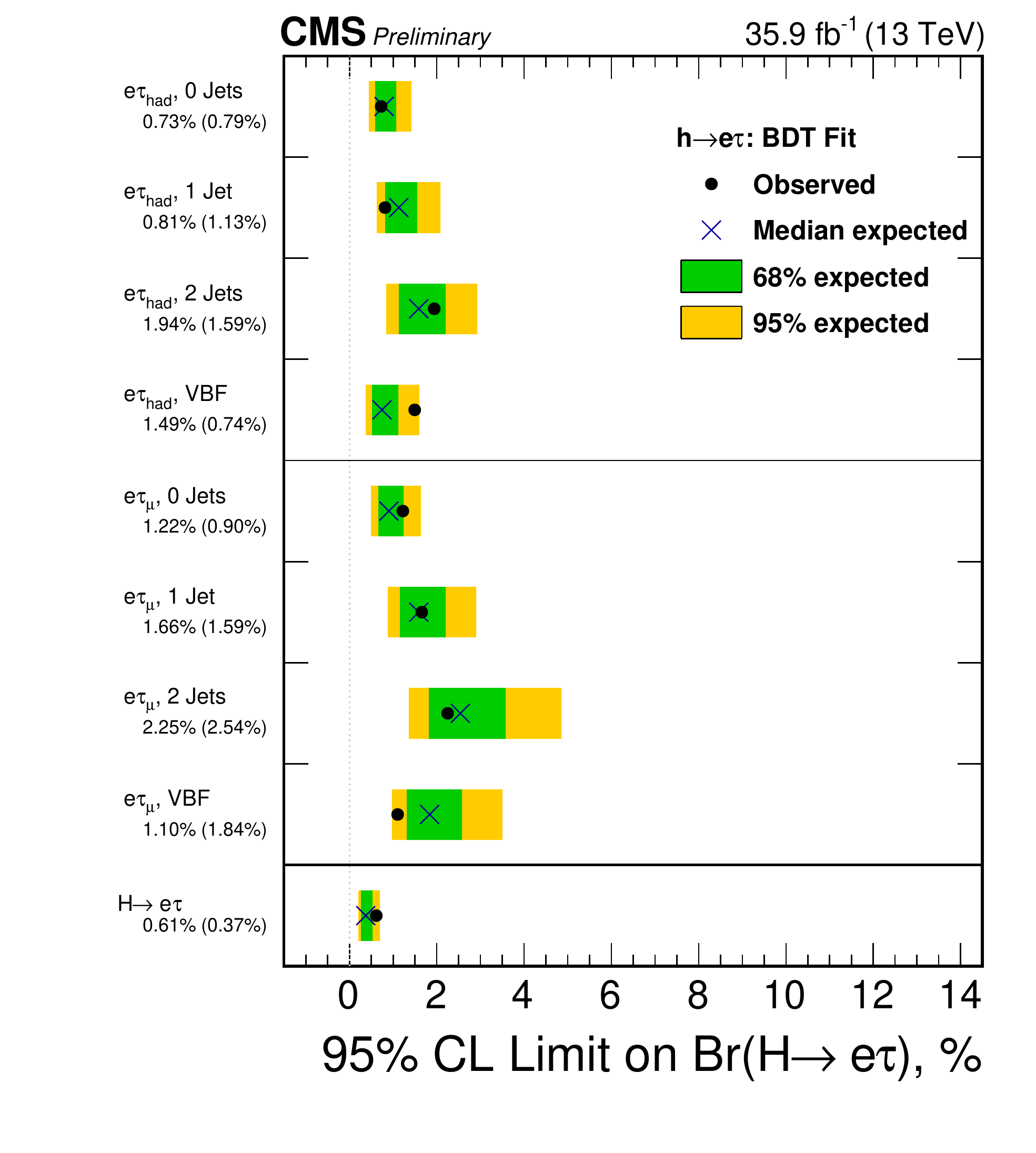}
\caption{ Upper limits on the branching ratios $\mathcal{B}(\text{H}~\rightarrow~\mathrm{e} \uptau)$ for different categories \cite{CMS_LFV_Run2}. }
\label{fig:LFV_CMS_limits}
\end{center}
\end{figure}

In the SM, lepton flavor violating (LFV) decays are forbidden since the Yukawa matrices are flavor diagonal in the lepton sector. However, lepton flavor decays are predicted in many BSM models, e.g. supersymmetric or composite Higgs models.~Analyzing LHC Run-1 data, CMS found a small excess of 2.4 standard deviations in the H $\rightarrow$ $\mu\uptau$ final state \cite{CMS_LFV_Run1}.~The corresponding ATLAS result can be found in Ref.~\cite{ATLAS_LFV_Run1}.~The updated CMS result using data collected at a center-of-mass energy of $\sqrt{\mathrm{s}}=13$TeV can be found in Ref.~\cite{CMS_LFV_Run2}.~The analysis considers 4 final states, namely the $\mu\uptau_\mathrm{h}$, $\mu\uptau_\mathrm{e}$, $\mathrm{e}\uptau_\mathrm{h}$ and $\mathrm{e}\uptau_\mathrm{\mu}$ states.~The signal is extracted from BDT distributions where the collinear mass is the key discriminating variable. No evidence for LFV Higgs boson decays are found.~Constraints on the non-diagonal elements of the Yukawa matrix are set  and upper limits are retrieved for the observed (expected) branching ratios:
\begin{itemize}
\item $\mathcal{B}(\text{H}~\rightarrow~\mu \uptau) < 0.25(0.25)\%$
\item $\mathcal{B}(\text{H}~\rightarrow~\mathrm{e} \uptau) < 0.67(0.37)\%$.	
\end{itemize}
Figure \ref{fig:LFV_CMS_limits} displays the obtained limits on the $\text{H}\rightarrow\mathrm{e} \uptau$ branching ratio.

\subsection{CP violating decays}

\begin{figure}[htb]
\begin{center}
\includegraphics[width=0.45\textwidth]{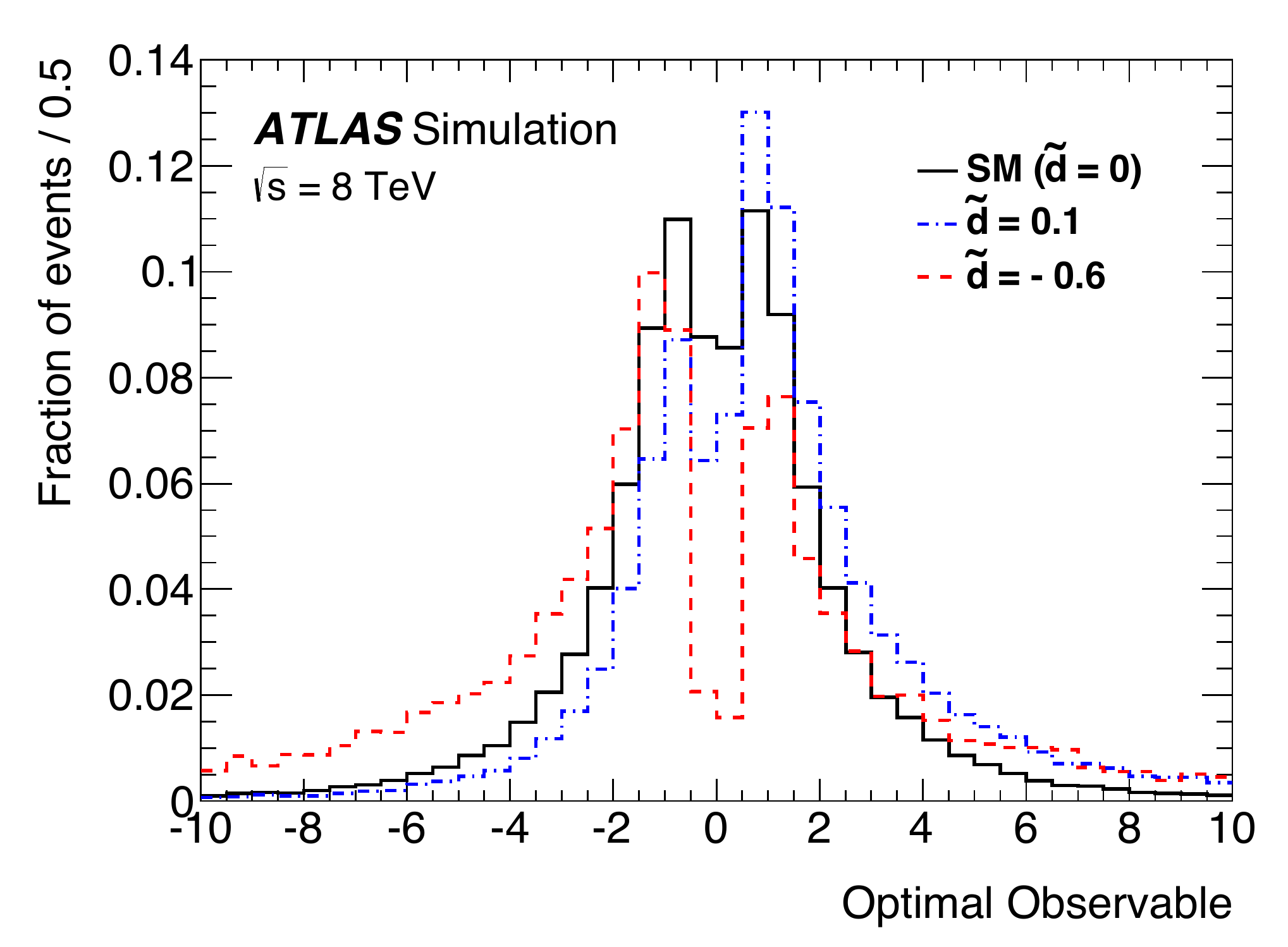}
\caption{  Distributions of the optimal observable $\mathcal{OO}$ for different values of the CP-mixing parameter $\tilde{d}$ \cite{ATLAS_CP}. }
\label{fig:ATLAS_CP}
\end{center}
\end{figure}

The search for CP violating interactions of the SM Higgs boson with other SM particles is motivated by the lack of CP violating sources to explain the baryon asymmetry that is observed in the universe.~In Ref.~\cite{ATLAS_CP}, the $\mathrm{H}\rightarrow\uptau\uptau$ channel is used to probe the couplings of a pair of vector bosons to the SM Higgs boson.~The analysis is based on 20.3 $\mathrm{fb^{-1}}$ of data collected at a center-of-mass energy of $\sqrt{\mathrm{s}}=8~$TeV. A CP-odd optimal observable $\mathcal{OO}$ is constructed:
\begin{equation}
	\mathcal{OO}=\frac{\mathrm{2Re(\mathcal{M}^*_{SM}\mathcal{M}^*_{CP-odd})}}{|\mathcal{M}_{\mathrm{SM}}|^2},
\end{equation}
where $\mathcal{M}_{\mathrm{SM}}$ and $\mathcal{M}_{\mathrm{CP-odd}}$ are SM CP-even and CP-odd contributions, respectively. An expectation value $<\mathcal{OO}\neq 0$ would be an indication of CP violation.~A CP-mixing parameter $\tilde{\mathrm{d}}$ is defined which relates the SM CP-even and CP-odd contributions:
\begin{equation}
	\mathcal{M}=\mathcal{M}_{\mathrm{SM}} + \tilde{\mathrm{d}} \cdot \mathcal{M}_{\mathrm{CP-odd}}.
\end{equation} 
Figure~\ref{fig:ATLAS_CP} shows the distributions of the optimal observable $\mathcal{OO}$ for different values of the CP-mixing parameter $\tilde{d}$.~No CP-violation is found and the analysis allows to constrain $\tilde{\mathrm{d}}$ in the interval $[-0.11,0.05]$ at $68\%$ confidence level, which is consistent with SM predictions of $\tilde{\mathrm{d}}=0$.

\section{High Luminosity LHC prospects of Higgs analyses}

In 2018, after LHC Run-2, the integrated luminosity will be $\approx 150~\mathrm{fb}^{-1}$.~In 2023, after LHC Run-3, the integrated luminosity is foreseen to be $\approx 300~\mathrm{fb}^{-1}$.~Finally, in 2037, after the High Luminosity LHC area, it is planned to have collected $\approx 3000~\mathrm{fb}^{-1}$ of data.~Table~II summarizes the presented results.~The respective ATLAS and CMS uncertainties are $\approx30\%$ and start to become statistically dominated.  

\begin{center}
\begin{table}[htbp]
\begin{tabular}{ | c | c | }
\hline
Analysis & Result \\ \hline \hline
$\displaystyle \mathrm{\mu_{CMS}^{\uptau\uptau}}$ & $ \displaystyle \mathrm{1.06^{+0.11}_{-0.09}(th.)\textcolor{red}{^{+0.13}_{-0.12}(syst.)}^{+0.12}_{-0.12}(bbb)^{+0.15}_{-0.15}(stat.)}$ \\ \hline
$\displaystyle \mathrm{\mu_{CMS}^{bb}}$ & $\displaystyle \mathrm{1.19^{+0.21}_{-0.20}(stat.)\textcolor{red}{^{+0.34}_{-0.32}(syst.)}}$ \\ \hline
$\displaystyle \mathrm{\mu_{ATLAS}^{bb}}$ & $\displaystyle \mathrm{1.20^{+0.24}_{-0.23}(stat.)\textcolor{red}{^{+0.34}_{-0.28}(syst.)}}$ \\ \hline
$\displaystyle \mathrm{\mu_{ATLAS}^{ttH}}$ & $\displaystyle \mathrm{1.20^{+0.2}_{-0.2}(stat.)\textcolor{red}{^{+0.3}_{-0.2}(syst.)}}$ \\

\hline
\hline
\end{tabular}
\label{tab:summary}
\caption{ Summary of the presented results.~A special emphasis in put on the systematic uncertainties.}
\end{table}
\end{center}

\begin{figure}[htbp]
\begin{center}
\includegraphics[width=0.4\textwidth]{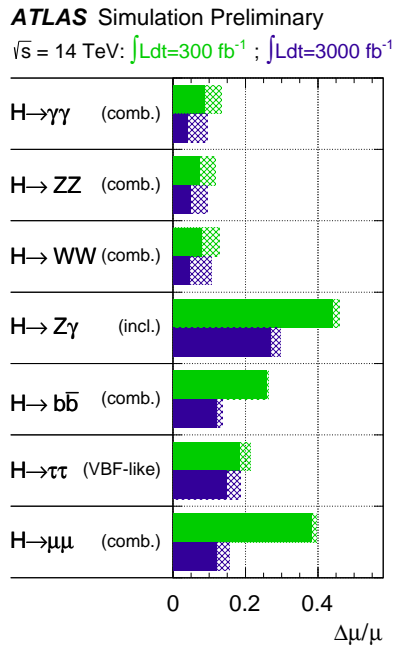}
\caption{ Relative uncertainty on the signal strength $\mu$ with $300~\mathrm{fb}^{-1}$ and $3000~\mathrm{fb}^{-1}$ of 14 TeV LHC data for various Higgs boson final states where a SM Higgs boson is assumed with a mass of 125 GeV \cite{ATLAS_pros}.  }
\label{fig:ATLAS_projection}
\end{center}
\end{figure}

\begin{figure}[htbp]
\begin{center}
\includegraphics[width=0.45\textwidth]{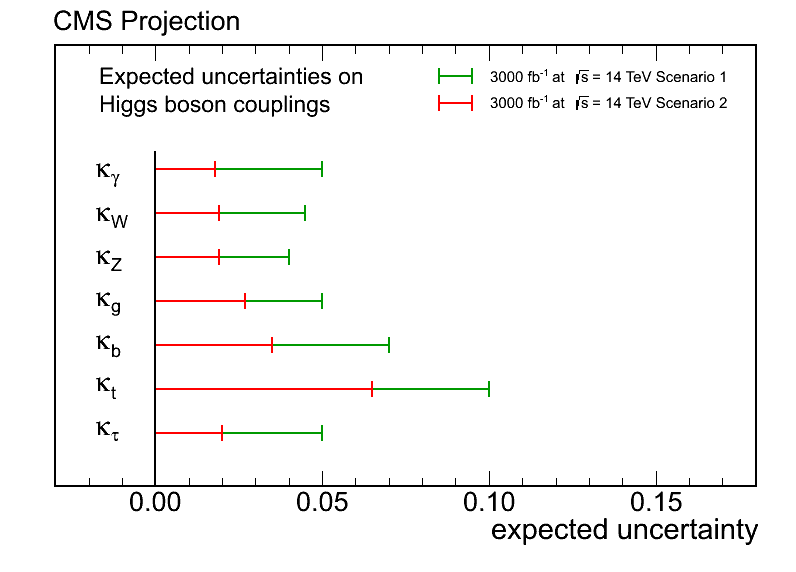}
\caption{  Estimated precision on the measurements for modified couplings for a SM Higgs boson assuming $300~\mathrm{fb}^{-1}$ and $3000~\mathrm{fb}^{-1}$ of 14 TeV LHC data.~The projections are obtained with two uncertainty scenarios: The green scenario where all systematic uncertainties are left unchanged and the red scenario where the theoretical uncertainties are reduced by a factor of two and all other uncertainties are reduced by the square root of the integrated luminosity \cite{CMS_pros}. }
\label{fig:CMS_projection}
\end{center}
\end{figure}

The goal for upcoming analyses is to decrease the systematic uncertainties as much as possible where it is to consider that several systematic uncertainties partially scale with the integrated luminosity.~After the High Luminosity LHC area, the Higgs boson relative couplings could be known with a precision of up to a few percent.~Figure~\ref{fig:ATLAS_projection} summarizes the relative uncertainty on the signal strength by ATLAS \cite{ATLAS_pros} and Fig.~\ref{fig:CMS_projection} displays the estimated precision on the modified couplings by CMS \cite{CMS_pros}.

\newpage

\section{Summary}

The latest ATLAS and CMS results and Higgs boson couplings to fermions are presented. The decay of the Higgs boson into a pair of tau leptons is the first fermionic decay to be observed by CMS with a significance of more than five standard deviations.~Individual searches for a Higgs boson decaying into a pair of bottom quarks show significances for more than three standard deviations both for ATLAS and CMS.~Searches for Higgs boson production in association with a pair of bottom quarks are performed in various decay channels.~So far all results agree with a SM Higgs boson within uncertainties.~The presented results are summarized and the systematic uncertainties are compared.Projections to an integrated luminosity of  $3000~\mathrm{fb}^{-1}$ show that after the High Luminosity LHC area, the relative couplings could be known with a precision of up to a few percent.

\end{document}